\documentclass[fleqn,usenatbib]{mnras}

\usepackage{newtxtext,newtxmath}

\usepackage[T1]{fontenc}

\DeclareRobustCommand{\VAN}[3]{#2}
\let\VANthebibliography\thebibliography
\def\thebibliography{\DeclareRobustCommand{\VAN}[3]{##3}\VANthebibliography}

\usepackage{graphicx}	
\usepackage{amsmath}	

\title[Precession of broken Be star discs]{Fast nodal precession of the disc around Pleione requires a broken disc}

\author[R. G. Martin \& S. Lepp]{
Rebecca G. Martin,\thanks{E-mail: rebecca.martin@unlv.edu}
and Stephen Lepp
\\
Nevada Center for Astrophysics, University of Nevada, Las Vegas,
4505 South Maryland Parkway, Las Vegas, NV 89154, USA\\
Department of Physics and Astronomy, University of Nevada, Las Vegas,
4505 South Maryland Parkway, Las Vegas, NV 89154, USA\\
}

\date{Accepted XXX. Received YYY; in original form ZZZ}

\pubyear{2022}

\begin{document}
\label{firstpage}
\pagerange{\pageref{firstpage}--\pageref{lastpage}}
\maketitle

\begin{abstract}
Pleione is a Be star that is in a 218 day orbit with a low-mass binary companion. Recent numerical simulations have shown that a Be star disc can be subject to breaking when material is actively  being  fed into the inner parts of the disc. After breaking, the disc is composed of two rings: an inner ring that is anchored to the stellar equator and an outer ring that is free to nodally precess.   A double ring disc may explain some of the observed variability in Pleione.  
We model the nodal precession of the outer disc ring that is driven by the companion on an observed timescale of $80.5\,\rm yr$. 
We find that the outer ring of a broken disc in a binary with an eccentricity of $e_{\rm b}= 0.6$ can precess on the observed timescale and have an outer radius that is in rough agreement with the observed disc size. 
An unbroken disc model cannot fit both the observed precession rate and disc size. Suppression of Kozai-Lidov driven disc eccentricity is more likely for a high binary eccentricity if the disc extends to the tidal truncation radius.  
\end{abstract}




\section{Introduction}

Be type stars are rapidly rotating stars \citep{Slettebak1982,Porter1996} that have shown H$\alpha$ emission at least once \citep{Porter2003}. They have a viscous Keplerian decretion disc \citep{Pringle1991} that forms from the ejection of material from the stellar equator \citep{Lee1991,Hanuschik1996,Carciofi2011}.  While accretion is actively occurring from the Be star to the disc, the inner disc feels a torque from the addition of material to the disc. If the spin of the Be star is misaligned to the binary orbital plane, then the torque from the companion drives nodal precession \citep{Lubow2000,Ogilvie2001b,martin2011be}.  As a result of these competing torques, the disc may be subject to breaking if it is not in sufficiently good radial communication \citep{Nixonetal2013,Dogan2015,Suffak2022}.

A disc can precess like a solid body when the communication timescale is shorter than the precession timescale \citep{Papaloizou1995,Larwoodetal1996}. The communication in the disc may be through viscosity (if the disc aspect ratio is smaller than the \cite{SS1973} viscosity parameter, $H/R \ll \alpha$) or through pressure (if $H/R \gg \alpha$) \citep[e.g.][]{Nixon2016}. 
Be star discs may be flared, meaning that $H/R$ increases with radius, while the viscosity is large, $\alpha \approx 0.1-0.3$ \citep{Jones2008,Rimulo2018,Ghoreyshi2018,Martin2019,Granada2021}. Disc breaking has been seen in 3D hydrodynamical simulations in the pressure dominated bending wave regime \citep[e.g.][]{Martin2018} and in the viscous regime \citep[e.g.][]{Nixonetal2013}. The outer parts of Be star discs may be in the intermediate regime where $\alpha \approx H/R$ where the communication may be a combination of viscosity and pressure \citep[e.g.][]{Martin2019warps}. Recent 3D hydrodynamical simulations of Be star disc simulations have shown breaking. The individual rings are in good communication across their radial extent and they are not significantly warped \citep{Suffak2022}.

Observations of the variability in Pleione may be explained by a disc comprised of two misaligned rings \citep{Tanaka2007}. In Section~\ref{obs} we first explore the properties of the Pleione system and what the precessing disc model tells us about the binary properties. Motivated by recent modelling of the precession of the Pleione system \citep{Marr2022},  we then explore a physically motivated precession timescale for the outer ring of a broken disc. 
In Section~\ref{model} we consider the orbits of misaligned test particles in the system and then use the precession timescale of these to examine an analytical model for the precession rate of the outer disc ring.  We  constrain the radial extent of the outer disc ring by the observed precession timescale and draw our conclusions in Section~\ref{conc}.

\section{Observational constraints on Pleione}
\label{obs}

The best fitting mass for Pleione is $M_1=4.1\,\rm M_\odot$ \citep{Marr2022}. 
Pleione  has a binary companion, although its orbital properties are not well determined. The orbital period is $218\,\rm days$ and the eccentricity may be $e_{\rm b}=0.6$ \citep{Katahira1996,Katahira1996b} or even as high as almost $e_{\rm b}=0.8$ \citep{Nemrovova2010}. However, there remains the possibility that the orbit is close to circular and the measured eccentricity is a result of phase-dependent effects of the circumstellar material \citep{Nemrovova2010}. In this work we consider three values for the binary eccentricity, $e_{\rm b}=0$, 0.6 and 0.8. 

The mass function is observed to be $f(M)=(M_2 \sin i_{\rm los})^3/(M_1+M_2)^2=2.4\times 10^{-4}\,\rm M_\odot$, where $i_{\rm los}$ is the inclination of the disc to the line of sight with an inclination of $0^\circ$ corresponding to a face on disc. 
Since there is no detected X-ray emission, the companion is unlikely to be a neutron star or a black hole. The most likely secondary is low mass dwarf, later than A5 V with mass less than Pleione, although a helium star or a white dwarf companion remain possible \citep{Katahira1996}.

Observations suggest that the outer disc ring around Pleione  nodally precesses on a timescale of $80.5 \,\rm yr$.  The inclination between the disc ring and the binary angular momentum vectors is $59^\circ$ and the precession axis is inclined by $84^\circ$ to the line of sight \citep{Hirata2007,Marr2022}. The disc ring mass is low compared with the binary mass and so the precession axis for the disc ring should coincide with the binary angular momentum vector, assuming that any alignment towards the binary orbital plane takes place on a longer timescale than the precession.  With $i_{\rm los}=84^\circ$ we find the companion mass is  $M_2=0.37\,\rm M_\odot$. This is the value that we take in the rest of our work.

The density of the Be star disc scales with the radius from the Be star, $R$, as $\rho \propto R^{-n}$ where $n=2.7$ \citep{Marr2022}. We consider this value and also a larger value of $n=3.5$ in this work. \cite{Marr2022} found the outer disc radius to be approximately $66\,\rm R_\odot$. This is in rough agreement with \cite{Hirata2007} who found the disc to extend up to a varying radius between $10\,R_1$ up to $25\,R_1$. With their value for $R_1=3.67\,\rm R_\odot$, this is the approximate range $37\,\rm R_{\odot}$ up to $92\,\rm R_\odot$.

\section{Broken disc ring precession timescale}
\label{model}

\begin{figure}
\begin{center}
\includegraphics[width=1\columnwidth]{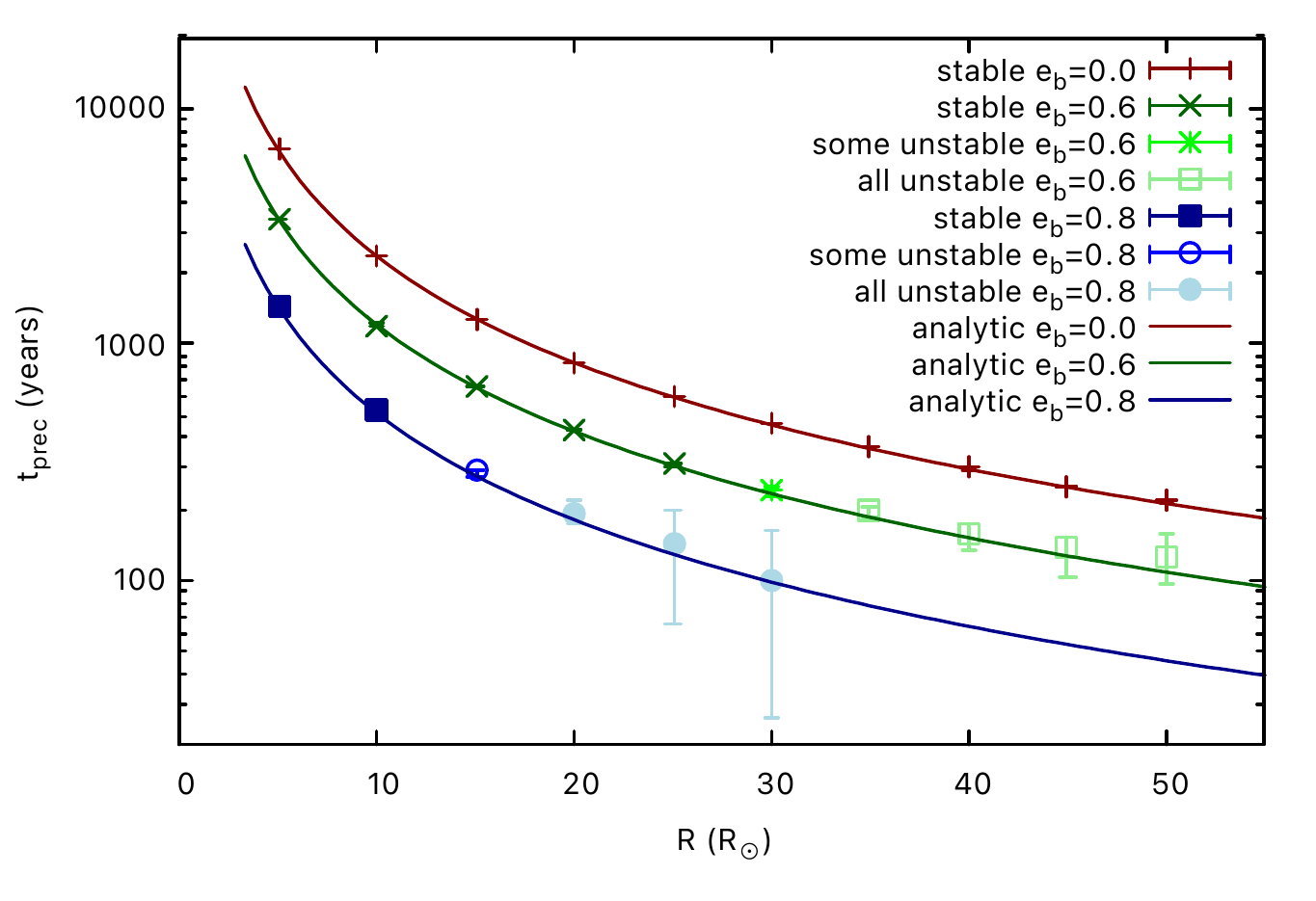}
	\end{center}
    \caption{Nodal precession timescale as a function of the particle semi-major axis for a binary system with $M_1=4.1\,\rm M_\odot$, $M_2=0.37\,\rm M_\odot$, orbital period $P_{\rm orb}=218\,\rm day$ and initial particle inclination $i=60^\circ$. Each point represents the average of 25 numerical simulations with different initial true anomaly and longitude of ascending node. The error bars show the range of the results. The analytic lines are equation~(\ref{eqprec}). 
    }
    \label{testparticles}
\end{figure}

 In this Section we first consider the evolution of misaligned test particle orbits around the Be star and then we use the precession timescale of these to calculate properties of the disc.

\subsection{Particle precession timescale}

The binary star has semi-major axis $a_{\rm b}$. The particle has semi-major axis of $R$ around mass $M_1$ and Keplerian angular frequency $\Omega=\sqrt{GM_1/R^3}$. The particle orbit is initially misaligned by inclination $i$ to the binary orbital plane. 
The particle undergoes nodal precession   on a timescale of
\begin{equation}
    t_{\rm prec}=\frac{2\pi}{\omega},
    \label{eqprec}
\end{equation}
where the precession frequency is
\begin{equation}
    \omega_{\rm particle}=\frac{3}{4} \frac{M_2}{\sqrt{M_1 M}}\frac{\Omega_{\rm b}}{(1-e_{\rm b}^2)^{3/2}}\left(\frac{R}{a_{\rm b}}\right)^{3/2}\cos(i)
    \label{omega}
\end{equation}
\citep{Bateetal2000,Okazaki2017}, the total mass of the binary is $M=M_1+M_2$ and the binary angular frequency is $\Omega_{\rm b}=\sqrt{GM/a_{\rm b}^3}$.

We use the $n$-body code {\sc rebound} \citep{Rein2012} to numerically integrate the orbits of test particles that are misaligned by $i=60^\circ$ to the binary orbital plane for three different binary eccentricities and varying orbital radius of the particle.  For each radius and eccentricity we run a five by five grid of evenly spaced initial true anomalies and initial longitudes of the ascending nodes.  We  calculate the initial precession rate of the particles over the first eight binary orbital periods. If any of the orbits  reaches an eccentricity of one before 16 binary orbits we do not calculate a precession rate.  During this time, averaged over a binary period, the particles precess at a constant rate. We note that at this inclination of $i=60^\circ$, the particles are unstable to Kozai-Lidov oscillations of eccentricity and inclination \citep{Kozai1962,Lidov1962}.  During periods of high eccentricity,  the precession is not uniform in time. However, since we are ultimately interested in how a disc behaves, we consider the initial precession rate during the time of low eccentricity. We discuss in Section~\ref{KL} the disc parameters for which the KL oscillations may be suppressed in the disc.

The results of the simulations are shown in Fig.~\ref{testparticles} along with the analytic fit given by equation~(\ref{eqprec}). There is good agreement between the numerical solutions and the analytic approximation.  
The numerical points show only the stable orbits. The larger the eccentricity of the binary, the smaller the radius at which the particles become unstable \citep[e.g.][]{Holman1999}. We define a particle to be unstable when its orbit reaches an eccentricity of one in less than 50,000 binary orbits \citep[e.g.][]{Quarles2018,Chen2020}.   We have identified radii as "stable" if none of the simulations go unstable in this time, as "some unstable" if less than 25 go unstable and as "all unstable" if all 25 initial conditions go unstable.
We note that even though some orbits later become unstable, their initial precession timescale is in good agreement with the analytic prediction.

\begin{table}
\begin{center}
\begin{tabular}{l c c c c c}
\hline
 $e_{\rm b}$ &  $R_{\rm L1}/{\rm R_\odot}$ & $n$ &  $R_{\rm out, ring}/{\rm R_\odot}$ & $R_{\rm out, no break}/{\rm R_\odot}$ \\
 \hline
\hline
0 & 183  & 2.7 & 95 & 158 \\
  &        & 3.5 & 95 & 209 \\
  \hline
0.6 &  73   & 2.7 & 61 & 101 \\
&        &  3.5 &  61 & 132 \\
\hline
0.8 &  36 & 2.7 & 34 & 56  \\
&       &3.5 & 34 & 71  \\
\hline
\end{tabular}
\end{center}
\caption{ The first column is the binary eccentricity. The second column  shows the distance of Pleione to the Roche lobe radius of the secondary at periastron separation, which we take to be a strict maximum for the radius of the disc. The third column shows the density exponent. The fourth and fifth columns show the minimum and maximum outer disc radii respectively for a disc ring that precesses on a timescale $t_{\rm prec}=80.5\,\rm yr$. If $R_{\rm out,no break} > R_{\rm L1}$ then the solution is unphysical and there is no possible unbroken solution.} 
\label{table2}
\end{table}

\subsection{Disc ring precession timescale}

Misaligned rings of a disc around the Be star feel the same tidal torque as the particles. However, the rings in the disc are connected to each other and instead the disc can precess on roughly the same timescale at all radii if the communication is faster than the precession timescale.  We note that the outer radius of the disc may be larger than the outermost stable particle orbit found in the previous section because the communication through the disc from the inner parts is able to stabilise the disc farther out \citep[e.g.][]{Martin2022}.

We assume that a fully extended disc first undergoes disc breaking to form two disc rings \citep[e.g.][]{Suffak2022}. The inner ring is anchored to the Be star equator while the outer disc is free to nodally precess.  We assume that the outer disc ring is in good radial communication across its radial extent and precesses rigidly, close to a solid body, as in the simulations presented in \cite{Suffak2022}.
We use an analytic approximation to the disc precession timescale  rather than running costly hydrodynamic simulations since this allows us to efficiently  consider a wide range of parameter space.  To calculate the precession rate of the disc we use
\begin{equation}
     \omega_{\rm disc}=\frac{3}{4} \frac{M_2}{\sqrt{M_1 M}}\frac{\Omega_{\rm b}}{(1-e_{\rm b}^2)^{3/2}}\left<\left(\frac{R}{a_{\rm b}}\right)^{3/2}\right>\cos(i),
\end{equation}
where the bracketed term involves the angular momentum weighted average over the disc given by
\begin{equation}
    \left<\left(\frac{R}{a_{\rm b}}\right)^{3/2}\right> = 
    \frac{\int_{R_{\rm in}}^{R_{\rm out}}\Sigma R^3 \Omega (R/a_{\rm b})^{3/2}\,dR}
    {\int_{R_{\rm in}}^{R_{\rm out}}\Sigma R^3 \Omega\,dR}
\end{equation}
\citep[see e.g.][]{Papaloizou1995,Larwoodetal1996,Lubow2018} and $\Sigma$ is the disc surface density. The use of this formula assumes that the disc ring itself is in good radial communication and remains relatively flat. In Section~\ref{conc} we discuss the possibility for strong warping.

The inner radius of the outer disc ring is equal to the break radius in the disc, $R_{\rm in}=R_{\rm break}$, with the minimum possible inner radius being the stellar radius that we take to be $3.67\,\rm R_\odot$ (with this minimum possible radius the disc is not broken).  The maximum possible outer radius is the tidal truncation radius although this is more difficult to determine. This was found by \cite{Paczynski1977} for a cold disc. For $e_{\rm b}=0$, this radius is around $137\,\rm R_\odot$. The outer disc edge  scales with $1-e_{\rm b}$ \citep{Artymowicz1994} and so with $e_{\rm b}=0.6$, this is about $55\,\rm R_\odot$ and for $e_{\rm b}=0.8$ this is around $27\,\rm R_\odot$.  However, we note that the tidal truncation radius may be much larger than this for hot and inclined discs \citep{Papaloizou1977,Papaloizou1983, Lubowetal2015,Miranda2015,Brown2019,Cyr2017}. The tidal torque falls off strongly with the inclination as $\cos^8(i)$ \citep{Lubowetal2015}.  Since the tidal truncation radius is not easily determined without hydrodynamical simulations, we consider the strict maximum value for the outer disc ring radius as the radius of the Lagrange L1 point from Pleione
that is shown in Table~\ref{table2}.  We calculate this by computing the difference between the periastron separation of the binary and the average Roche lobe radius for the secondary with equation 2 in \cite{Eggleton1983}. We do this rather than use the Roche lobe radius of the Be star because the disc may be larger than its Roche lobe for high inclination \citep[e.g.][]{Miranda2015}.

\begin{figure*}
\begin{center}
\includegraphics[width=1.6\columnwidth]{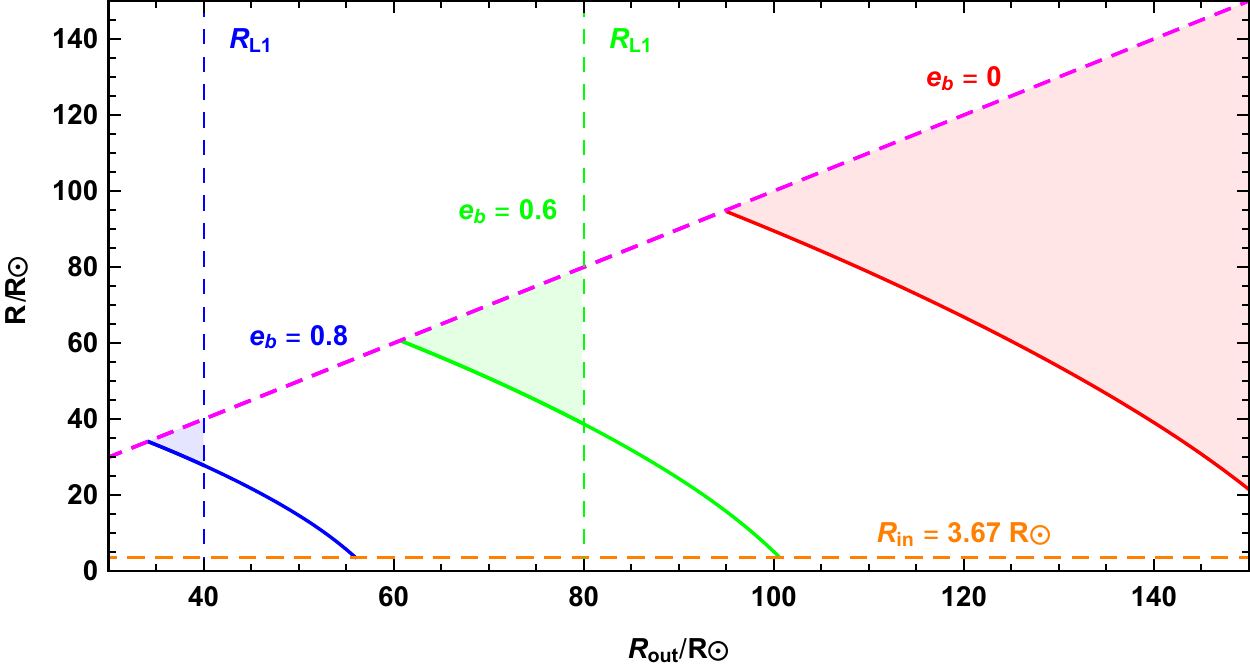}
	\end{center}
    \caption{Radial extent of a disc ring with density $\rho \propto R^{-n}$ with $n=2.7$ with a nodal precession rate of $t_{\rm prec}=80.5\,\rm yr$ for a circular orbit binary (red), eccentric binary with $e_{\rm b}=0.6$ (green) and $e_{\rm b}=0.8$ (blue). The solid lines show the inner radius of the outer ring, $R_{\rm in}$, for a given outer disc radius, $R_{\rm out}$.  The vertical length of the shaded regions show the radial extent of the outer disc ring. The dashed orange line shows the stellar radius and the magenta dashed line shows $R_{\rm out}$. The disc, by definition, can only be in $R<R_{\rm out}$, meaning under the magenta line. The vertical dashed lines show the minimum location of the Roche lobe radius of the secondary (there is no line for the circular case since it's off the scale).} 
    \label{trunc2}
\end{figure*}
 We assume that the disc is isothermal \citep[e.g.][]{Carciofi2011} and the aspect ratio scales as $H/R\propto R^{1/2}$. The surface density $\Sigma=H\rho$ then scales as $\Sigma \propto R^{3/2-n}\propto R^{-1.2}$ for $n=2.7$. Since the disc mass is low compared to the binary, the absolute value of the density does not affect the dynamical evolution, only the distribution of the material.

We now consider the radial extent of an outer disc ring that precesses on the observed rate. Fig.~\ref{trunc2} shows the radial extent of a disc ring with the observed precession timescale $t_{\rm prec}=80.5\,\rm yr$ for a range of outer disc truncation radii for the three different binary eccentricities. In Table~\ref{table2} we also show the minimum and maximum possible outer disc radius for each model for a disc ring that precesses on the observed timescale. If the outer disc radius is equal to the minimum radius, $R_{\rm out}=R_{\rm out,ring}$, then the solution is a narrow ring at that radius. If the outer disc radius is equal to the maximum possible radius, $R_{\rm out}=R_{\rm out,no break}$, then the solution is a disc that extends all the way down to the stellar radius, an unbroken disc. For both of  the eccentric orbit cases,  $R_{\rm out,no break}>R_{\rm L1}$, and therefore an unbroken disc is not a possible solution. The disc cannot extend down to the stellar surface and precess at the observed rate. Therefore the only solutions are for a broken disc. In the circular orbit case, a disc that extends down to the surface is possible, but the size of such a disc of $R_{\rm out}=210\,\rm R_\odot$ would be difficult to reconcile with the observed disc size. Therefore, there are no possible unbroken disc solutions that match the observed precession rate and the observed disc size.

We have also considered a density exponent $n=3.5$ and the minimum and maximum values for the outer disc radius are shown in Table~\ref{table2}. The minimum outer radius is not sensitive to $n$, while the maximum outer radius increases with $n$. 

The outer disc radius of $66\,\rm R_\odot$ observed by \cite{Marr2022} is in agreement with a narrow outer disc ring precessing in the $e_{\rm b}=0.6$ case.  A higher binary eccentricity leads to a disc that is too small to be reconciled with the observed disc size, while the circular binary disc is too large.
\subsection{Disc eccentricity growth}
\label{KL}

Such a highly misaligned disc, can be unstable to KL disc oscillations \citep{Martinetal2014, Fu2015b}.  The timescale for KL oscillations is approximately the same as the nodal precession timescale \citep{Ford2000,Kiseleva1998}. However, the KL oscillations are very sensitive to the initial disc conditions and the time for the first cycle may be delayed while the disc remains circular \citep[e.g.][]{Fu2015}. KL disc oscillations have been suggested to operate in Be/X-ray binaries since a large disc eccentricity may cause significant mass transfer to the companion and therefore the observed type~II outbursts \citep[e.g.][]{Martinetal2014b,Franchini2019, Franchini2021}. The binaries that show these outbursts typically have shorter orbital periods than that of Pleione \citep{Cheng2014}.

\begin{figure}
\begin{center}
\includegraphics[width=0.8\columnwidth]{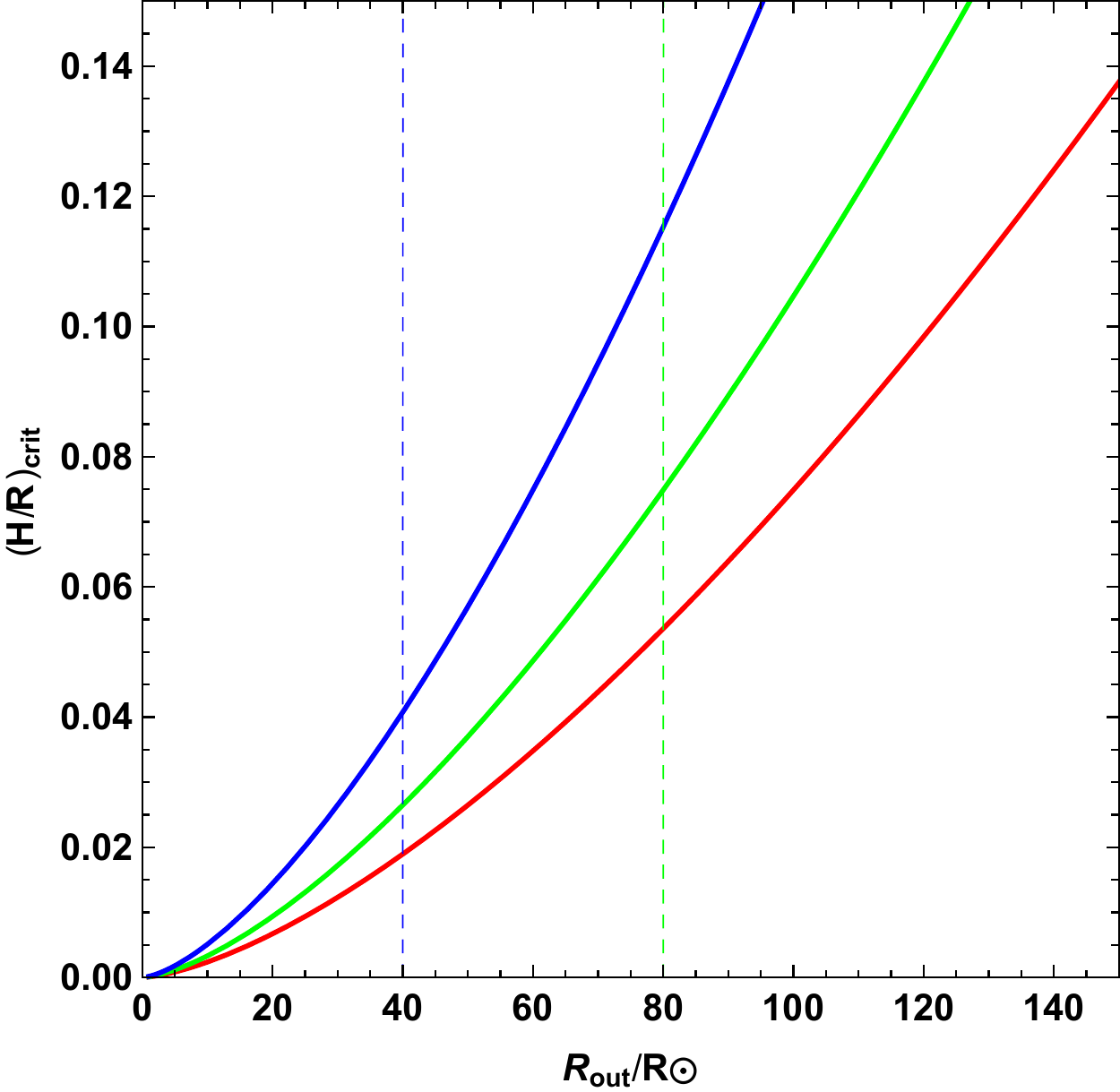}
	\end{center}
    \caption{The critical disc aspect ratio above which the disc KL is suppressed for $e_{\rm b}=0$ (red), $e_{\rm b}=0.6$ (green) and $e_{\rm b}=0.8$ (blue). The vertical dashed lines show the maximum possible disc size, $R_{L1}$, as the periastron separation of the binary minus the Roche lobe radius of the secondary. }
    \label{kl}
\end{figure}

The KL disc instability may be suppressed for a sufficiently large disc aspect ratio \citep{Lubow2017,Zanazzi2017}. This is because the pressure induced apsidal precession rate exceeds the binary induced rate \citep{Martinetal2014}. Since the Be star disc is flared with radius, the instability is more likely to be suppressed for longer orbital period binaries \citep{Martin2019kl}. 
The disc KL oscillations are suppressed if the aspect ratio at the outer edge is greater than the critical value that is given by
\begin{equation}
    \left(\frac{H}{R}\right)_{\rm crit}\approx \sqrt{\frac{M_2 }{M_1}}\sqrt{\frac{R^3}{a_{\rm b}^3}}(1-e_{\rm b}^2)^{-3/4}
\end{equation}
\citep{Lubow2017},
where we have included the eccentricity dependence of the precession rate. Since the outer disc radius scales with $(1-e_{\rm b})$, the critical disc aspect ratio at the outer disc edge scales as $((1-e_{\rm b})/(1+e_{\rm b}))^{3/4} $, which decreases with eccentricity.  The form of the eccentricity dependence suggests that KL is more easily suppressed in Be star discs in highly eccentric orbit binaries (assuming that the disc extends to its tidal truncation radius). Fig.~\ref{kl} shows the critical disc aspect ratio as a function of the disc outer radius. For the eccentric binaries, the critical aspect ratio is low even at the maximum disc size, $R_{\rm L1}$, and these discs are more likely to be stable against KL. For the circular orbit binary, the critical aspect ratio at the outer disc edge may be much larger because the disc is larger. Therefore suppression of the KL oscillations in the circular orbit binary is less likely. Three-dimensional numerical hydrodynamic simulations are required in order to explore this process more thoroughly for the specific binary orbital parameters of this system \citep[see][]{Suffak2022}.

\section{Discussion and conclusions}
\label{conc}

Motivated by the recent observations and conceptual model of a broken disk  around the Be star Pleione presented by \cite{Marr2022}, we have considered a physically motivated model of the precessing decretion disc ring. We  included the effects of the binary companion that truncates the outer edge of the disc and drives nodal precession.  We found that in order to explain both the observed disc size and the nodal precession rate,  the binary eccentricity must be around $e_{\rm b}\approx 0.6$ and the  disc must be broken. The outer ring of a disc that is broken has a faster precession rate than a disc that extends down to the stellar surface.

A double ring disc structure may form through the tearing of an initially continuous disc \citep[e.g.][]{Nixonetal2013}. Broken and misaligned disc rings have recently been observed in the protoplanetary disc around GW Ori \citep{Bi2020,Kraus2020}. The cause of the break in this case may be a result of the inner binary torque, or the presence of gap-forming planets.  In the case of the Be star disc, the broken disc is a result of the competing torques on the disc: the accretion torque and the binary companion torque.  The viscous timescale in the disc is relatively short. For example, if we take $H/R\propto R^{1/2}$ and $H/R=0.04$ at $R=3\,\rm R_\odot$, then the viscous timescale at $R=100\,\rm R_\odot$ is only $1.5\,\rm yr$ with $\alpha=0.3$ \citep[e.g.][]{Pringle1981}. Therefore, an outer ring with an inner radius larger than the stellar radius must be prevented from rapidly spreading inwards by the presence of the inner ring.

Since the disc around Pleione is observed to be misaligned by about $60^\circ$ to the binary orbital plane, KL disc oscillations are possible. However, they are more easily  suppressed with a high binary eccentricity because of the smaller disc tidal truncation radius.  We have shown that if the aspect ratio $H/R\gtrsim 0.08$ at the disc outer edge with $e_{\rm b}=0.6$, then the disc KL oscillations are suppressed.
If a disc is KL unstable, the precession rate (and the inclination of the disc relative to the binary) may vary in time. The average precession rate increases when the disc is KL unstable.  \cite{Suffak2022} found that the addition of material at the disc inner edge suppressed the KL disc oscillations. The disc is KL unstable once decretion from the Be star to the disc has ended.

If the disc was strongly warped, rather than broken, then communication may not be cut off from the outer disc to the inner disc. Since the addition of material to the inner parts of the disc is always aligned to the stellar equator, the torque from the mass addition can only act to slow the overall precession rate of the disc. The precession period could be longer than the freely precessing model that we have used here \citep{Okazaki2017}. This therefore suggests that a warped disc that extends down to the stellar radius is unable to precess fast enough to explain the observed rate.

\section*{Acknowledgements}
We thank an anonymous referee for useful comments that improved the manuscript. We acknowledge support from NASA through grant 80NSSC21K0395.

\section*{Data Availability}

 The data underlying this article will be shared on reasonable request to the corresponding author.
 



\bibliographystyle{mnras}
\bibliography{martin} 








\bsp	
\label{lastpage}
\end{document}